\documentclass[twocolumn,switch]{article} 
\usepackage[utf8]{inputenc}
\usepackage{enumerate}
\usepackage{lipsum}
\usepackage{calc}
\usepackage[T1]{fontenc}
\usepackage{csquotes}
\usepackage[hyphens]{url}
\usepackage{microtype}
\usepackage{nicefrac}
\usepackage{textcomp}
\usepackage{lipsum}
\usepackage{tabularx}
\usepackage{flushend}
\usepackage{xspace}
\usepackage{ifthen}
\usepackage{multirow}
\usepackage{float}
\usepackage{longtable}
\usepackage{array}
\usepackage{booktabs}
\usepackage{graphicx}
\usepackage{enumitem}
\usepackage{wrapfig}
\usepackage{rotating}
\usepackage{setspace}
\usepackage{bigstrut}
\usepackage{verbatim}
\usepackage{wasysym}
\usepackage{soul}
\usepackage{vcell}
\usepackage{pifont}
\usepackage{afterpage}
\usepackage{newfloat}
\usepackage{sidecap}
\usepackage{subcaption}
\usepackage[font=footnotesize]{caption}
\usepackage{titlesec}
\usepackage{circledsteps}
\usepackage{tikz}
\usepackage{xcolor}
\PassOptionsToPackage{colorlinks=true,linkcolor=purple,urlcolor=blue,citecolor=cyan,anchorcolor=black}{hyperref}
\usepackage{orcidlink}
\usepackage{hyperref}
\usepackage{eso-pic}
\usepackage{authblk}
\usepackage[numbers,square,sort]{natbib}
\usepackage[ruled,vlined]{algorithm2e}
\newcommand{\etal}[1][]{%
\ifthenelse{\equal{#1}{}}{et~al.\xspace}{et~al.~\cite{#1}\xspace}%
}

\usepackage{listings}

\definecolor{keywordcolor}{rgb}{0.0, 0.0, 0.6}
\definecolor{commentcolor}{rgb}{0.0, 0.5, 0.0}
\definecolor{stringcolor}{rgb}{0.6, 0.0, 0.0}
\definecolor{backgroundcolor}{rgb}{0.95, 0.95, 0.95}

\lstdefinestyle{LangiumReqStyle}{
    backgroundcolor=\color{backgroundcolor},
    basicstyle=\ttfamily\footnotesize,
    keywordstyle=\color{keywordcolor},
    commentstyle=\color{commentcolor}\itshape,
    stringstyle=\color{stringcolor},
    numbers=left,
    numberstyle=\tiny,
    numbersep=5pt,
    showspaces=false,
    showstringspaces=false,
    showtabs=false,
    tabsize=1,
    captionpos=b,
    breaklines=true,
    frame=single,
    xleftmargin=8pt,
}

\newcommand{\citesec}[1]{Section~\ref{sec:#1}}
\newcommand{\citefig}[1]{Fig.~\ref{fig:#1}}

\newcommand{\bullitem}[1]{\textbf{#1}}

\newcommand{\ttformat}[1]{\texttt{\small {#1}}}

\usepackage[mathlines, switch]{lineno} 

\DeclareFloatingEnvironment[name={Supplementary Figure}]{suppfigure}
\sidecaptionvpos{figure}{c}

\usepackage{amsmath,amssymb,amsfonts}

\newcommand{\Description}[1]{}
\newcommand{\textcite}[1]{\cite{#1}}

\titlespacing\section{0pt}{12pt plus 3pt minus 3pt}{1pt plus 1pt minus 1pt}
\titlespacing\subsection{0pt}{10pt plus 3pt minus 3pt}{1pt plus 1pt minus 1pt}
\titlespacing\subsubsection{0pt}{8pt plus 3pt minus 3pt}{1pt plus 1pt minus 1pt}

\newcolumntype{L}[1]{>{\raggedright\let\newline\\\arraybackslash}p{#1}}
\newcolumntype{C}[1]{>{\centering\let\newline\\\arraybackslash}p{#1}}
\newcolumntype{M}[1]{>{\centering\arraybackslash}m{#1}}
\newcolumntype{R}[1]{>{\raggedleft\let\newline\\\arraybackslash}p{#1}}

\definecolor{alizarin}{rgb}{0.82, 0.1, 0.26}

\newcommand{\agenthi}[1]{\texttt{\small #1}}

\usepackage{xcolor}
\usepackage{mdframed}
\usepackage{caption} 
\usepackage{tikz}
\definecolor{promptgray}{rgb}{0.96, 0.96, 0.96}

\newmdenv[
    linecolor=black,
    linewidth=0.5pt,
    backgroundcolor=promptgray,
    innerleftmargin=8pt,
    innerrightmargin=8pt,
    innertopmargin=6pt,
    innerbottommargin=6pt,
    roundcorner=0pt, 
    skipabove=0pt,
    skipbelow=0pt
]{promptframe}

\definecolor{myorange}{HTML}{FFFFFF}
\definecolor{reqblue}{HTML}{b3d9ff}
\definecolor{secred}{HTML}{ffb3b3}

\newcommand{\fwcomp}[1]{\ttformat{#1\xspace}}

\newcommand{\reqcir}[1]{%
  \tikz[baseline=(char.base)]{%
    \node[shape=circle,draw,inner sep=0pt,fill=reqblue,minimum size=1em] (char) {\fwcomp{#1}};%
  }%
}
\newcommand{\seccir}[1]{%
  \tikz[baseline=(char.base)]{%
    \node[shape=circle,draw,inner sep=0pt,fill=secred,minimum size=1em] (char) {\fwcomp{#1}};%
  }%
}

\newcommand{\repcir}[1]{%
  \tikz[baseline=(char.base)]{%
    \node[
      shape=circle,
      draw,
      inner sep=0pt,
      minimum size=1em,
      path picture={
        \fill[reqblue]
          (path picture bounding box.west) rectangle
          (path picture bounding box.north east);
        \fill[secred]
          (path picture bounding box.south west) rectangle
          (path picture bounding box.east);
      }
    ] (char) {\fwcomp{#1}};
  }%
}

\usepackage{preprint}
\AddToShipoutPictureBG*{%
  \ifnum\value{page}=1
    \put(0,20){%
      \makebox[\paperwidth]{\hfill \small\ttfamily\shortstack{Accepted for publication at the 34th IEEE International Requirements Engineering Conference (RE@Next'26).\\*correspondence: Zoe.Pfister@uibk.ac.at} \hfill}%
    }%
  \fi
}

\title{Transforming Privacy Artifacts into Accessible Reports for Non-Technical Stakeholders}

\author[1\thanks{\tt{Zoe.Pfister@uibk.ac.at}}]{Zoe~Pfister \orcidlink{0009-0009-2882-5059}}
\author[1]{Clemens~Sauerwein \orcidlink{0009-0009-9464-5080}}
\author[1]{Benedikt~Dornauer \orcidlink{0000-0002-7713-4686}}
\author[2]{Tina~Mersch \orcidlink{0009-0000-3284-3318}}
\author[2]{Christian~Wolf \orcidlink{0009-0009-9377-0699}}
\author[1]{Ruth~Breu}
\author[1]{Michael~Vierhauser \orcidlink{0000-0003-2672-9230}}

\affil[1]{University of Innsbruck\\Department of Computer Science\\Austria}
\affil[2]{EKS InTec\\Germany}

\begin{document}

\twocolumn[ 
  \begin{@twocolumnfalse} 

\maketitle   

\begin{abstract}
The transition toward Industry 5.0 is reshaping industrial work environments with an emphasis on human-centricity, enabling close collaboration between humans and machines to enhance productivity and flexibility.
However, such systems typically require monitoring of human workers and operators, often involving sensitive data, raising significant privacy concerns.
As a result, affected workers and unions frequently reject human-machine collaboration features due to a lack of transparency regarding privacy threats and implemented mitigation strategies.
To enable early stakeholder involvement, establish trust, and support informed decision-making, privacy implications must be communicated in a way understandable to non-technical stakeholders. 
Yet, current Requirements Engineering (RE) practices provide limited methodological support for making privacy threats and mitigations accessible to non-technical stakeholders (e.g., individual workers or their representative unions). 
In this paper, we propose a conceptual framework that guides software design from human monitoring-related use cases and requirements to informed decision-making guidance focusing on non-technical stakeholders.
Building on principles such as Privacy by Design, the framework leverages Large Language Models (LLMs) to transform technical artifacts into accessible privacy reports. We share initial insights from two industry use cases, evaluate the quality of the generated reports, and outline future research directions toward integrating privacy transparency into RE processes for human-centric industrial systems.
\end{abstract}
\vspace{0.35cm}

  \end{@twocolumnfalse} 
] 


\section{Introduction}\label{sec:intro}

The transition toward Industry 5.0 emphasizes human-centricity~\cite{leng_industry_2022} and promotes close collaboration between humans and machines. In particular, these interaction points require stakeholder-centered Cyber-Physical Systems (CPS) engineering and operation~\cite{michailidis2021continuous}. 
The overarching objective is to enhance productivity while simultaneously increasing flexibility~\cite{kopp_success_2021}. To ensure correct and safe operation, as well as enable self-adaptation during collaborative tasks, systems must implement monitors to collect and analyse diverse runtime information~\cite{zheng2016efficient,stojadinovic2015dynamic,vierhauser2018monitoring}. 
In scenarios involving human collaboration, monitored data is not solely limited to pure machine-related data, but also includes information about human operators and workers~\cite{buerkle2022adaptive,liu2017ar}. For example, consider a mobile robot carrying heavy tooling while following a worker~\cite{siemensrobots}. To operate safely, the robot must track both the operator and nearby humans to prevent collisions.
We refer to requirements that involve the collection and processing of data about human stakeholder activities as \emph{human monitoring requirements}.

\ieeecopyright

However, monitoring humans at runtime inevitably raises the issue of collecting potentially sensitive information, posing significant privacy concerns, potentially hampering, or even preventing the acceptance and implementation of collaborative features altogether.
In the past, several cases of extensive workplace surveillance and misuse have resulted in pushback, public backlash, and regulatory scrutiny. For example, Amazon France Logistique was fined 32 million euros in 2023 for intrusive employee monitoring that violated GDPR principles~\cite{amazon_bad_2023}.
In a similar case in 2020, H\&M was fined 35 million euros for monitoring several hundred employees at a service centre in Germany~\cite{hm_bad_2020}.
These and many other examples highlight how privacy concerns can hinder the acceptance of monitoring-enabled systems and fuel opposition from workers and unions.

From a requirements perspective, human monitoring requirements often face significant criticism and resistance from worker unions, employee representatives, and privacy advocates. Monitoring practices are frequently regarded as intrusive and ethically problematic, with workers and their representatives viewing such technologies with suspicion due to concerns about privacy violations, job quality, reduced autonomy, and employee dignity~\cite{singh2019much,oz1999electronic}.
In many cases, collective bodies and unions frequently argue that such systems should not be implemented without transparent safeguards, informed consent, clear documentation of purpose, and effective mitigation strategies to address privacy risks~\cite{eurofund2020,connolly2012dataveillance}.
In addition, compliance with regulatory frameworks, such as the GDPR, introduces further constraints during system design.
While several researchers have investigated how the GDPR and similar regulations can be integrated and aligned with Requirements Engineering (RE) processes or how privacy-related requirements can be extracted  ~\cite{kosenkov2025privacy,negri2024understanding,abualhaija2025llm,manasreh2025designing}, current RE practices still lack structured processes that bridge privacy design artifacts and informed decision-making by non-technical stakeholders.
As a result, privacy threats, trade-offs, and mitigation strategies must be communicated in a way accessible to non-technical stakeholders, including workers and their representatives, to increase transparency and support technology acceptance.

In this paper, we address this challenge by integrating privacy transparency considerations into Software Engineering (SE) and RE processes. We develop a first iteration of a framework that guides software design from human monitoring requirements to informed decision-making guidance by workers and their representatives. 
Specifically, we employ established security and privacy analysis techniques, such as Data Flow Diagrams (DFD) and the STRIDE framework~\cite{conklin_threat_2025}, and provide these artifacts as input to Large Language Model (LLM) agents. 
More specifically, several agents transform the technical artifacts -- together with the use cases and requirements --  into accessible privacy reports. 
With this approach, we intend to increase involvement and trust of non-technical stakeholders.

In this early research, we aim to pursue the following research objectives (ROs): 
\begin{enumerate}[
    align=left, 
    leftmargin=!,
    labelwidth=\widthof{\textbf{RO1:}}, 
]
    \item[\textbf{RO1:}] {Develop a structured RE process that supports privacy analysis and communication of privacy threats and mitigation tactics related to human monitoring requirements, thereby facilitating broader understanding and technology acceptance among stakeholders.}
    \item[\textbf{RO2:}] {Investigate how LLM-assisted transformations of technical privacy artifacts can improve the accessibility and understandability of privacy threats and mitigation strategies for non-technical stakeholders.}
\end{enumerate}

\begin{figure*}[tb!]
    \centering
    \includegraphics[width=\textwidth]{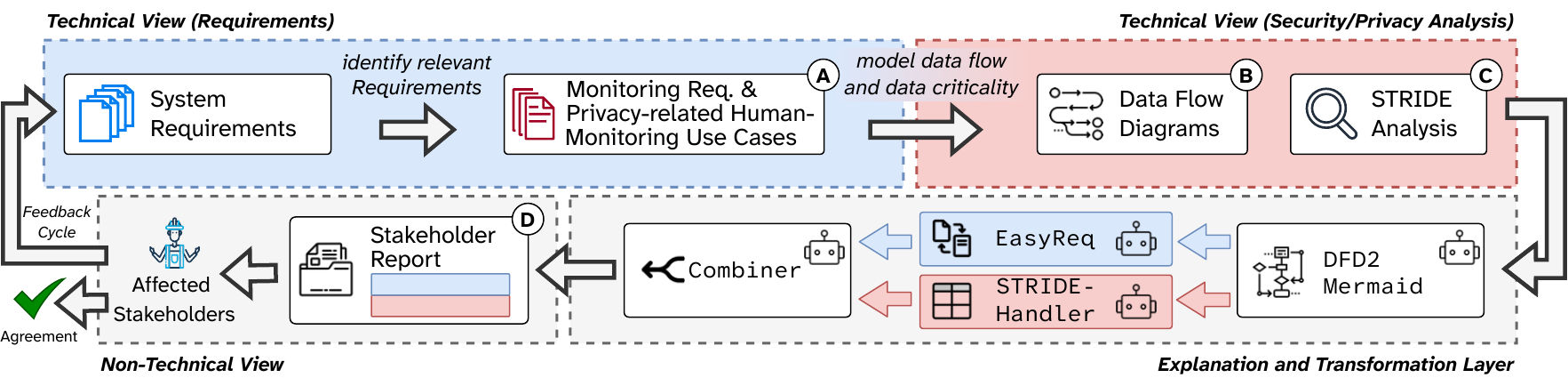}
    
    \caption{ Overview of our proposed framework, transforming human monitoring requirements and privacy analysis artifacts into stakeholder-oriented privacy reports.}
    \label{fig:concept-diagram}
    \vspace{-.6cm}
\end{figure*}

Our contributions in this paper are threefold: First, we present a structured process that links human monitoring requirements, privacy design artifacts, and use cases to decision-making processes in worker unions.
Second, we propose an approach to increase technology acceptance in Industry 5.0 by making privacy threats and their mitigations more accessible through LLM-assisted reporting. This includes generating human-readable reports that explain the privacy implications of human monitoring requirements and planned mitigations. Third, we report on a preliminary validation, gaining insights through a questionnaire and feedback session with our industry collaborator, and outline directions for future research areas.

The remainder of this paper is laid out as follows: \citesec{background} presents background, existing research, and our motivating example. In \citesec{framework}, we introduce our framework and process steps. We apply the framework using the motivating example in \citesec{implementation}. \citesec{experiment} presents a proof-of-concept validation using two industry use cases. Finally, Sections~\ref{sec:discussion-and-roadmap} and \ref{sec:conclusion} outline our research roadmap and conclusions.\\\vspace{-0.8em}

\textbf{Data availability:} We provide all supplementary material as part of a repository:~\url{https://github.com/Ethical-Human-Machine-Interaction/PrivacyArtifacts2Report}.
\section{Background \& Related Work}
\label{sec:background}

Privacy in industrial CPS concerns how workers’ personal data is collected, processed, and shared.  In monitoring-enabled environments, such data may include behavioural, positional, or performance-related information about workers. Regulatory frameworks such as the GDPR treat privacy as a design concern by requiring data protection measures and transparent, accessible information for data subjects~\cite{gdpr_regulation_2016}.

\textbf{Security and Privacy Analysis Techniques:} In Software Engineering, Data Flow Diagrams are commonly used to visualize how data moves through a system by modelling processes, data stores, external entities, data flows, and trust boundaries~\cite{sion_solutionaware_2018}. DFDs also form the basis for STRIDE-based threat modelling. STRIDE is a widely used approach for systematically identifying security threats during system design~\cite{hernan2006threat,shostack2014threat}. It categorises threats into six classes: \textbf{S}poofing, \textbf{T}ampering, \textbf{R}epudiation, \textbf{I}nformation Disclosure, \textbf{D}enial of Service, and \textbf{E}levation of Privilege. STRIDE analyses each element of a DFD for potential threats within these categories. This structured approach helps developers identify security risks early in the design phase and derive appropriate mitigations before implementation. Since STRIDE explicitly addresses information disclosure threats, which are particularly relevant in privacy-sensitive systems~\cite{hernan2006threat}, DFDs combined with STRIDE provide a link between GDPR-oriented privacy analysis and structured threat identification.

\textbf{LLMs/Agents supporting RE Activities:} LLMs are transformer models~\cite{vaswani_attention_2017} trained on large text corpora and designed to generate plausible continuations of textual prompts~\cite{wu_ai_2022}.  
One key aspect is their ability to perform \enquote{in-context learning}, i.e., adopting new tasks at runtime based on natural language instructions (prompts)~\cite{wu_ai_2022}.
This flexibility has led to increasing interest in using LLMs to support various activities in the SE lifecycle~\cite{krishna_using_2024,vogelsang_specifications_2024,hou_large_2024a}. However, LLMs often act as black boxes, lacking transparency (e.g., through hallucinations) and controllability, especially in complex tasks~\cite{huang_survey_2025}. 
This limitation is particularly relevant when generating privacy-related artifacts for stakeholders, where each artifact must be rigorously validated and, if necessary, edited by a human to ensure accuracy and correctness.
To address this problem, Wu et al.~\cite{wu_ai_2022} propose \emph{Chaining LLM steps together}, where a complex problem is decomposed into multiple sub-problems executed sequentially. Each step receives inputs from previous steps of the chain. Naturally, humans can analyse the intermediate outputs and edit them should they detect incorrect statements.

Another important factor influencing output quality is prompt formulation. For example, Brown et al.~\cite{brown_language_2020} have shown that few-shot prompting, where the prompt includes examples of the task to be completed, can significantly improve model performance compared to zero-shot prompts.
A more recent approach is \emph{chain-of-thought prompting}~\cite{wei_chainofthought_2022}, where the input prompt contains chain-of-thought examples of the task to solve.
This increased performance on arithmetic and complex symbolic reasoning tasks, among others.
Additional work has also explored template-based prompting approaches for structuring LLM interactions in specific domains~\cite{tian_templatebased_2025}.

To better understand what monitoring data is collected, and how privacy may be affected, we enrich traditional use case templates (e.g., goals, scenarios) with monitoring aspects. The  \emph{``monitoring use cases''}~\cite{value_2025_pfister} describe what data is collected during monitoring, what equipment is used for collection, and who the stakeholders being monitored are. 
These monitoring use cases can then be further enriched with information for a GDPR Data Privacy Impact Assessment (DPIA)~\cite{henriksen-bulmer_dpia_2020,gdpr_regulation_2016}.


\textbf{Motivating Example:} Our motivating example is drawn from a use case defined by our industry collaborators EKS InTec (\emph{InLine Control of Product Assembly}). The goal is to improve product quality during assembly.
To achieve this, the system uses camera-based tracking to monitor the assembly process and identify deviations or faulty components.
When such deviations are detected, the system notifies shop-floor workers in real time, temporarily pausing the assembly process until the error is corrected.  While the primary functionality of the use case revolves around quality assurance in the assembly process, it necessitates continuous monitoring of workers. This introduces several privacy concerns related to the collection of video footage and behavioural information of shop-floor workers. Without systematic and rigorous privacy modelling and the translation of these technical specifications into transparent privacy reports, the system risks both non-compliance with GDPR and opposition from worker unions due to the sensitive nature of the captured video and behavioural data.



\section{Towards a Privacy Transparency Framework for Human Monitoring Use Cases}
\label{sec:framework}
Our framework (cf.~\citefig{concept-diagram}) builds upon established software engineering artifacts. System engineers identify use cases and requirements that involve monitoring of human stakeholders. These \reqcir{A} \emph{human monitoring requirements} and \emph{use cases} serve as the starting point for subsequent privacy analysis and stakeholder communication activities.
In the following, we describe the framework and its core components.

\subsection{Framework Overview}
After identifying relevant monitoring use cases and requirements, the framework applies established security and privacy analysis techniques. Specifically, this involves constructing \seccir{B}  a Data Flow Diagram and \seccir{C} performing a STRIDE analysis. These artifacts support the systematic identification of potential privacy risks affecting stakeholders and the development of planned mitigation strategies to address them.
Together with the human monitoring requirements and use cases, these artifacts can also support the preparation of a GDPR-mandated Data Protection Impact Assessment (DPIA)~\cite{gdpr_regulation_2016}.
For example, the artifacts align well with the \emph{DPIA Data Wheel}~\cite{henriksen-bulmer_dpia_2020}, which maps questions about data processing activities to GDPR principles, facilitating regulatory compliance.

The use case, associated requirements, Data Flow Diagram, and STRIDE analysis then serve as input to the LLM Agents within the \emph{Explanation and Transformation Layer}. 
Within this layer, the complex technical artifacts are transformed into a \repcir{D} \emph{privacy report tailored to non-technical audiences}.
The resulting reports are presented to stakeholders involved in the decision-making process of accepting or rejecting the proposed privacy-related requirements (e.g., shop-floor workers or worker unions). The goal is to facilitate informed stakeholder discussions and support consensus-building regarding the acceptability of the proposed solutions. 
Based on feedback, the process either results in formal agreement on privacy requirements or triggers an iterative feedback cycle in which the use case and requirements can be revised and adapted.


\subsection{Explanation and Transformation Layer}\label{sec:explanation-transformation}

The core component of the framework is the \emph{Explanation and Transformation Layer}, which bridges the gap between technical artifacts and the perspectives of non-technical stakeholders. It receives the monitoring use case, its associated requirements, DFD, and the STRIDE analysis, and transforms them into a structured privacy report. This transformation is performed by a workflow of four specialized LLM agents:

First, the \agenthi{DFD2Mermaid Agent} converts a DFD image into a machine-readable Mermaid\footnote{See \url{https://mermaid.ai}.} diagram and generates a textual summary describing its components, data flows, and trust boundaries. This serves two purposes: reducing the number of tokens required to process the graph, and providing a human-readable representation of the model, allowing a human reviewer to validate the LLM’s interpretation of the data flow and correct errors early in the process.

The conversion of technical artifacts into stakeholder-oriented explanations is handled by the \agenthi{EasyReq Agent} and the \agenthi{STRIDE-Handler Agent}. 
The latter receives the DFD Mermaid diagram and summary of the \agenthi{DFD2Mermaid Agent}, along with the original requirement and the STRIDE analysis, as input. 
It iterates over the STRIDE entries, summarizing each identified threat in language comprehensible to non-technical stakeholders.
The agent also enriches the summary with concrete privacy threat examples relevant to the given use case and explains how each identified threat is intended to be mitigated. 
In parallel, the \agenthi{EasyReq Agent} rewrites the original use case and requirements into a simplified version suitable for non-technical stakeholders. Further, it generates a short rationale explaining the purpose, intended benefit, and implementation motivation of each requirement (cf. \citefig{input-output-example}).

Both agents follow explicit output guidelines, as specified in their respective input prompts. 
They are explicitly instructed to avoid technical terminology and instead explain concepts using simple terms, while keeping the output concise. 
Further, the \agenthi{EasyReq Agent}  is instructed to focus on the \emph{what} and \emph{why} of the use case rather than the \emph{how}, while the \agenthi{STRIDE-Handler Agent} illustrates threats using concrete examples and addresses privacy and surveillance-related concerns. 
Finally, the \agenthi{Combiner Agent} combines the outputs of all preceding agents into a structured HTML privacy report.

\section{Prototype Implementation and Usage Example}\label{sec:implementation}

We implemented an initial research prototype of the Explanation and Transformation Layer workflow using \emph{n8n}~\cite{n8n_n8nio_2026}, a workflow automation platform for orchestrating agents. In future iterations, we plan to further extend and implement the workflow using LangChain\footnote{See \url{https://www.langchain.com}.} to allow for more flexible agent coordination and integration with additional LLMs.

The \agenthi{DFD2Mermaid Agent} uses Gemini 2.5 Pro, which consistently produced valid conversions from DFD images into Mermaid representations during preliminary experiments, in contrast to other models. The remaining agents use Claude Sonnet 4.5. Extended reasoning  (``Thinking Mode'') is only enabled for the \agenthi{STRIDE-Handler Agent}, where more complex reasoning is required to interpret and summarize STRIDE analysis results.
Prompts were initially designed manually using Claude Console following the RICE template (Role, Instructions, Context, Constraints, Examples)~\cite{vogelsang_specifications_2024}. 
We then iteratively refined these prompts using Claude Sonnet 4.5 to conform to the Claude Prompting Best Practices~\cite{anthropicpbc_improve_2024,anthropicpbc_prompting_2026}, e.g., structuring inputs via XML tags. Additionally, for both the \agenthi{EasyReq Agent} and \agenthi{DFD2Mermaid Agent}, we included a scratchpad section to the prompt, outlining a list of steps to follow during text generation (cf. Figure~\ref{fig:easyreq-prompt}), facilitating the model's chain-of-thought reasoning capabilities~\cite{wei_chainofthought_2022}.

\begin{figure}[htbp]
    \begin{promptframe}
        \small
        \sffamily
        \noindent You are a security engineer with expertise in translating technical security information for non-technical stakeholders. 
        
        Your task is to take a technical software security requirement and its use case, then rewrite the requirement in clear, accessible language that non-technical stakeholders, especially worker unions, can understand.
        
        You should also explain why this security requirement is important and planned.
        
        \vspace{0.4em}
        [input-documents]
        \vspace{0.4em}

        Your goal is to:
        \begin{enumerate}[leftmargin=1.5em, labelsep=0.5em]
            \item Rewrite the requirement in plain, non-technical language that avoids jargon, acronyms, and technical terminology
            \item Explain the value and rationale for why this requirement is planned
        \end{enumerate}

\vspace{0.4em}
        [guidelines for non-technical communication]
\vspace{0.4em}

        Before writing your final output, use the scratchpad to:
        \begin{enumerate}[leftmargin=1.5em, labelsep=0.5em]
            \item Identify the core purpose of the technical requirement
            \item Note any technical terms that need to be simplified or explained
            \item Think about what business risks this requirement mitigates
            \item Consider what stakeholders care about most
        \end{enumerate}

\vspace{0.4em}
        $<$scratchpad$>$[Your analysis here]$<$/scratchpad$>$
        
        [examples]
    \end{promptframe}
    \caption{Single-Shot Prompt Template with Chain-of-Thought Scratchpad used for Simplifying Requirements and Use Cases for Non-technical Stakeholders. Used by the \agenthi{EasyReq Agent}.}
    \label{fig:easyreq-prompt}
    \vspace{-.5cm}
\end{figure}

As this implementation serves as an initial end-to-end proof of concept, we intend to compare the current solution with other LLM providers, such as ChatGPT, or Le Chat\footnote{See \url{https://chatgpt.com} and \url{https://chat.mistral.ai/chat} respectively.} as part of future work (cf.~\citesec{discussion-and-roadmap}).

\begin{figure*}
    \centering
    \includegraphics[width=\textwidth]{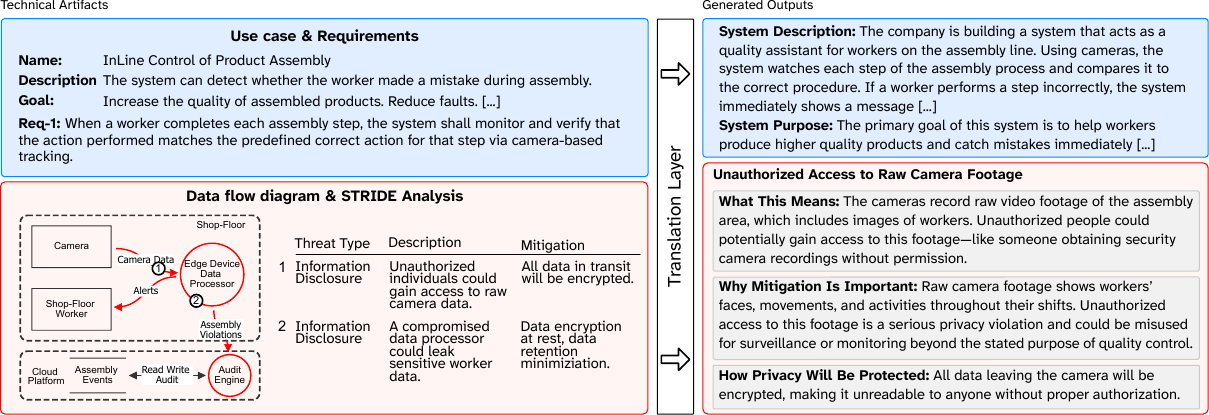}
    \caption{Example of Inputs and Generated Outputs of the Workflow based on the example (UC1) introduced in Section~\ref{sec:background}.}
    \label{fig:input-output-example}
    \vspace{-.3cm}
\end{figure*}

\textbf{Using the Framework:} We demonstrate our framework using the motivating example \emph{InLine Control of Product Assembly} introduced in \citesec{background}.
The workflow begins when a system engineer uploads design artifacts: \reqcir{A} the use case (including goals, scenarios, and monitoring properties) and associated requirements; \seccir{B} a Data Flow Diagram; and \seccir{C} the results of a STRIDE analysis.  A partial example is shown on the left side of \citefig{input-output-example}.
In this example, the DFD models how camera data flows from the sensor to an Edge Device Data Processor, which analyses the footage for assembly errors and forwards detected violations to a Cloud Platform (annotated as (1) in \citefig{input-output-example}).
Applying STRIDE to this data flow reveals a potential information disclosure threat, where unauthorized individuals could gain access to raw camera data if the data flow is insufficiently protected. A possible mitigation in this scenario is encryption.
Since the final report focuses on privacy aspects, we scoped the STRIDE analysis to the most relevant threat categories for privacy, namely \emph{Information Disclosure}, \emph{Spoofing}, and \emph{Tampering}. 

Based on these artifacts, the workflow transforms the data for non-technical stakeholders. The DFD is first converted to Mermaid syntax and summarized in natural language. The resulting output, together with the STRIDE analysis, the use case, and the requirements, is then provided as input to the \agenthi{STRIDE-Handler Agent}, which generates structured threat explanations. 
An example is shown on the bottom-right of \citefig{input-output-example}, where the information disclosure threat is converted into structured prose:

\begin{enumerate}[
    align=left, 
    leftmargin=!,
    labelwidth=\widthof{\textbf{1)}}, 
]
    \item \textit{Plain-language threat description (what):} The threat is described in accessible language, supported by illustrative examples for better clarification. In the example, the agent explains that unauthorized individuals could access raw camera footage.
    \item \textit{Explanation of stakeholder impact (why):} The potential consequences are described from the perspective of the stakeholders (e.g., shop-floor workers in the motivating example). The explanation highlights that unauthorized access to camera footage is a clear privacy violation and could be misused for surveillance beyond the intended purpose of quality control.
    \item \textit{Mitigation explanation (how):} The proposed mitigation strategy is discussed. In the example,  encryption is described as making the data unreadable to anyone without proper authorization.
    
\end{enumerate}

\noindent Each STRIDE threat is transformed following this structure.

Next, the \emph{EasyReq Agent} simplifies the original use case and requirements, producing a natural-language version and a rationale explaining the intended purpose/benefit of each requirement (cf. top-right side of \citefig{input-output-example}).

Finally, the \emph{Combiner Agent} integrates both outputs, generating the final report \repcir{D} in HTML format. The report begins with an executive summary, followed by a simplified system description and purpose statement produced by the EasyReq Agent. 
Next, the report lists the privacy threat explanations, each containing the three-part structure described above. 
This ordering ensures that a non-technical reader first understands what the system does and why it is being built before being introduced to potential privacy risks and mitigation strategies.

\section{Preliminary Validation}
\label{sec:experiment}
The goals of our approach are to develop a structured RE process that supports privacy analysis and communication to non-technical stakeholders (RO1) and to automatically generate an easily understandable privacy report based on a set of (technical) requirements and relevant standards and guidelines (RO2). Through this preliminary validation, we aim to obtain initial insights into the quality of our generated reports and the feasibility of our proposed process and workflow.
We, therefore, focus on two aspects: First, we demonstrate the feasibility of our proposed workflow using our end-to-end prototype applied to two real-world use cases (cf.~\citefig{input-output-example}) from our industry collaboration (cf.~\citesec{apply}). Second, we assess the quality of the generated reports using a combination of Software Requirements Specification (SRS) metrics and qualitative expert feedback through a survey and semi-structured interview with a domain expert from our industry collaborators (cf.~\citesec{feedback}).
The full set of interview questions and results can be found in the supplementary material.

\subsection{Validation Setup}

We were provided with two use cases by our industry collaborators: (UC1) \emph{InLine Control of Product Assembly}, monitoring workers to detect assembly errors and sequence violations, and (UC2) \emph{Cycle Time Monitoring for Lean Production}, tracking worker assembly durations to identify process inefficiencies. In both cases, requirements include camera-based monitoring of shop-floor workers, making them suitable candidates for privacy threat analysis and subsequent stakeholder-oriented reporting.

To assess report quality, we designed a questionnaire based on established document quality criteria. Since we are not aware of existing frameworks to evaluate non-technical privacy reports, we base our validation criteria on those proposed by Krishna et al.~\cite{krishna_using_2024}, who evaluate LLM-generated Software Requirements Specifications (SRS). 
Specifically, we collected six metrics: (1) internal consistency, (2) redundancy, (3) completeness with respect to technical artifacts, (4) conciseness, (5) correctness, and (6) understandability on 5-point Likert scales.
These criteria capture general aspects of document quality and provide a suitable baseline for evaluating structured privacy reports. After completing the questionnaire, the expert participated in a semi-structured interview to provide qualitative feedback on report structure and understandability.

The domain expert was provided with two reports: one manually created report -- created by one author, and reviewed by two additional authors -- and one automatically generated using our prototype implementation. The reports were presented in a blinded manner, i.e., the expert was not informed which report was generated by the framework. After answering the questions concerning the 6 quality metrics, the domain expert participated in a follow-up feedback session, where we conducted a semi-structured interview to gain more information on how to improve the report's structure and understandability in future work.

\subsection{End-To-End Report Generation}
\label{sec:apply}

For each of the two use cases and their requirements, we manually created a Data Flow Diagram and conducted a STRIDE analysis, focusing on the privacy-relevant threat categories of Information Disclosure, Spoofing, and Tampering.
Based on these artifacts, we generated reports using our workflow and manually created baseline reports for comparison.

Repeated executions of our workflow revealed minor inconsistencies in the generated output, confirming the non-deterministic nature of LLM-based workflows. For example, section titles for privacy violations were sometimes derived directly from threat titles of the STRIDE analysis, which might be unclear to non-technical stakeholders. We also observed minor formatting variations (bullet points vs. prose) across several workflow runs. To reduce visual bias, both manual and generated reports were formatted uniformly before being shared with our expert.

\subsection{Expert Feedback}
\label{sec:feedback}
\textbf{Quantitative Results:} Initial feedback from our expert indicates that the generated reports matched or outperformed the manually created baseline reports across all evaluated quality dimensions. The largest differences were observed in the completeness, conciseness, and understandability measures for UC1, where the generated report scored two points higher (on a 5-point scale). The internal consistency varied between the individual UCs, with the generated report scoring a 3 for UC1 and 5 for UC2. Related to understandability, the expert perceived the generated reports as more accessible from a language standpoint compared to the manually written ones. However, the generated reports introduced other issues, such as undefined abbreviations (e.g., LTPE for Lean Time Processing Engine), highlighting the need for human validation of generated outputs.

\textbf{Qualitative Results: } The follow-up interview provided additional insights into how the reports' understandability and structure could be further improved with respect to RO2.
Currently, each STRIDE threat is presented as a separate section in the report. However, the expert pointed out that this structure can result in perceived redundancy, making it difficult to understand that the sections actually describe different parts of the data flow.
For example, multiple threats in UC1 used encryption as their primary mitigation tactic  (e.g., camera to data processor to cloud platform).  Grouping such related threats and explaining them along the data flow could improve readability and reduce redundancy, making the report more concise.
The expert indicated that restructuring reports to reflect data flow dependencies and shared mitigation strategies complemented by a (simplified) visualization of the DFD (cf. Fig.~\ref{fig:input-output-example}), could further increase both conciseness and stakeholder comprehension. 
However, they highlighted that there is a trade-off, and care must be taken, ensuring that the report does not become too technical.
These findings provide initial evidence for RO2, while also highlighting areas for improvement in report structuring and presentation.

The expert further suggested enhancing the report with a more comprehensive introduction, describing the system in greater detail, as well as a concluding section summarizing key insights, open issues, and areas requiring stakeholder discussion. Additionally, introducing prioritization, e.g., using the MoSCoW principle (must, should, could, won't)~\cite{clegg_case_1994}, may serve as a basis for stakeholder deliberation. 


\subsection{Threats to Validity}

Our initial proof-of-concept demonstrates the feasibility and successful application of our framework and the generated reports. However, while the preliminary validation is based on a limited number of use cases, they represent real-world scenarios from our industry collaborators. 
To address this limitation, we are currently collaborating with industry partners to extend the set of use cases and apply the framework to more complex and diverse scenarios. 
In the initial prototype implementation and preliminary validation, we primarily used Claude Sonnet 4.5 (due to availability) for our agents. In future work, we plan to evaluate our framework using diverse LLMs (cf. \citesec{discussion-and-roadmap}). Despite this limitation, Claude Sonnet 4.5 produced promising results in the generated reports.

Finally, the interview revealed that inconsistencies within the input artifacts themselves propagate into the generated reports. For example, one artifact proposed encryption as a mitigation against threats where the adversary potentially holds the decryption keys. Since the LLM cannot produce internally consistent output from contradictory inputs, such issues carry over into the final reports. This highlights a fundamental limitation of the approach: the quality of the generated privacy report is bound by the quality and consistency of the initial technical artifacts, making rigorous validation of input documents a prerequisite for meaningful report generation.

\section{Discussion \& Research Roadmap}
\label{sec:discussion-and-roadmap}

Our initial validation, preliminary results, and expert feedback indicate that we are able to partially achieve our research objectives: our early version of the framework demonstrates how privacy analysis can be systematically integrated into early requirements engineering processes (RO1), and how structured, stakeholder-oriented privacy reports can be generated from technical artifacts (RO2).
Building on these results, we identify five research directions for our future work.

\bullitem{Integration of diverse Privacy Analysis Frameworks: } 
In our initial prototype, we employed the STRIDE framework~\cite{conklin_threat_2025} for systematically collecting and documenting privacy-relevant threats and corresponding mitigations. However, several alternative frameworks exist, and security and privacy researchers have developed various approaches for specifically analysing privacy risks, such as LINDDUN~\cite{wuyts2020linddun}, the DPIA Data Wheel framework~\cite{henriksen-bulmer_dpia_2020}, or STPA-Priv~\cite{shapiro_privacy_2016}. Given our modular architecture, these can be easily integrated as new agents.
As part of RO1 (cf. \citesec{intro}) -- extending our structured process -- we will analyse other assessment frameworks, investigate the strengths and limitations, and evaluate their suitability in early requirements elicitation processes to improve technology acceptance. 
Furthermore, we plan to develop an ontology, relating concepts across frameworks (threats, assets, controls, privacy goals). 
From this, we aim to derive recommendations and guidelines for selecting a framework based on project characteristics and the specific nature of human monitoring involved.

\bullitem{Mapping to Regulatory Texts:}  
Currently, input artifacts do not explicitly link identified threats and mitigations to relevant regulatory provisions, such as the GDPR~\cite{gdpr_regulation_2016}. This lack of traceability limits the applicability of the generated reports in compliance verification (e.g., GDPR Data Protection Impact Assessments). To address this limitation, we plan to enrich privacy threats and mitigation information with explicit mappings to relevant regulatory texts, including, for example, specific articles and recitals of the GDPR. This could be achieved through the integration of Retrieval Augmented Generation (RAG)~\cite{lewis_retrievalaugmented_2020} context pipelines supplying the agents with relevant regulatory passages when processing STRIDE artifacts, or through the integration of external knowledge sources via the Model Context Protocol (MCP)\footnote{See \url{https://modelcontextprotocol.io}.}. 
For example, mitigation tactics, such as data minimisation (cf. \citefig{input-output-example}), could be linked to Article 5(c) of the GDPR~\cite{gdpr_regulation_2016}. Such additional augmentations would enable the generation of specialized reports tailored to legal stakeholders and support compliance verification, further extending RO2.

\bullitem{Agent-Augmented Support:} 
Currently, our workflow relies on manually created design artifacts as input for report generation. In our ongoing work, we intend to extend the framework towards increased automation through LLM-supported artifact generation. This includes using custom agents that generate requirements from use cases, or conduct an automatic STRIDE analysis with tools such as STRIDE GPT\footnote{See \url{https://github.com/mrwadams/stride-gpt}.}. Another challenging topic for future work is the automated generation of DFDs from software requirements artifacts, where the LLM must make sound architectural decisions to satisfy all system requirements. While such automation has the potential to reduce modelling effort, it also shifts the burden towards validation.  This is especially critical for privacy and security aspects of SE, where incorrect threat assessments or missing mitigations may lead to non-compliance with the law, making rigorous human-in-the-loop validation essential. As part of this, we will investigate perceived accuracy and completeness of generated artifacts from the perspective of security professionals, as well as the associated cognitive load, for example, using established methods such as NASA-TLX.


\bullitem{Persona-specific Reports: } 
Stakeholders involved in the decision-making process differ substantially in their technical expertise and informational needs. For example, worker representatives may primarily be interested in understanding what data is collected and the privacy implications, whereas legal experts require detailed mappings to regulatory provisions to evaluate compliance. Currently, our framework generates a single report targeting non-technical stakeholders. Future work will introduce stakeholder persona-specific report generation to tailor reports to the specific concerns and expectations of different stakeholder groups, supporting RO2.
Persona-based customisation can be achieved by adapting LLM agent roles and prompt configurations. A key research challenge, therefore, is identifying which information is relevant for each persona, ensuring that simplification of the content does not compromise its completeness or correctness (cf. \citesec{experiment}).

\bullitem{Comprehensive evaluation:}
Our preliminary validation provides initial evidence that the LLM-generated privacy reports created via our framework accurately reflect relevant requirements and privacy analysis artifacts while improving accessibility to non-technical stakeholders.
However, a more comprehensive evaluation is required to fully address RO1 and RO2.
First, we plan to conduct an interview study with non-technical stakeholders and legal experts to better understand their expectations regarding privacy reports.  These insights will then inform the second iteration of our privacy report generation process.
Additionally, we plan to conduct a comparative evaluation of LLM-generated and manually created reports with both technical and non-technical stakeholders. 
Based on the feedback received from our expert interview, we will also explore automated validation methods (e.g., round-trip reconstruction of reports and technical artifacts, LLM-as-a-Judge~\cite{li2025generation}) to identify information loss or inconsistencies in generated reports and reduce manual validation efforts.










\section{Conclusion}
\label{sec:conclusion}
In this paper, we presented a Requirements Engineering framework that integrates privacy threat analysis with automated, non-technical stakeholder-oriented report generation. By combining established security analysis techniques, such as STRIDE and Data Flow Diagrams, with a modular LLM-based transformation pipeline, the approach enables the systematic derivation of structured, non-technical privacy reports from technical Software and Requirements Engineering artifacts. 
Our preliminary validation demonstrates both the feasibility of the approach and its potential in future work. At the same time, the results highlight important challenges, including the consistency of input artifacts, the need for improved report structuring, and the importance of human validation in the loop.
Overall, the findings suggest that LLM-supported transformations can serve as an effective mechanism to operationalize privacy-by-design principles within requirements engineering processes. As part of our research roadmap and ongoing work, we will extend the framework to incorporate additional privacy analysis methods, support artifact creation with LLMs, and conduct an empirical study to further assess the effectiveness and generalizability of our framework.
\section*{Acknowledgment}
This work was supported by the ITEA 4 project GENIUS, with funding from the Austrian Research Promotion Agency (FFG; grants 931318 and 921454) and the German Federal Ministry of Research, Technology and Space (BMFTR; grant 16IS24069D).

\small
\bibliography{citations-cleaned}

%

\end{document}